\documentclass[12pt]{article}
	
\usepackage{amsmath}
\usepackage{psfrag,graphicx,verbatim,array,multicol,palatino,enumerate,float}
\usepackage{amsmath,amsfonts,bm,rotating,natbib}
\usepackage{pstricks,pst-node,setspace}
\usepackage{epsfig}
\usepackage{longtable}
\usepackage{algorithm}
\usepackage[noend]{algpseudocode}
\usepackage[footnotesize,bf]{caption}

\def\bbeta{\pmb{\beta}}

\def\bmu{\pmb{\mu}}

\def\btheta{\pmb{\theta}}

\def\bSigma{\pmb{\Sigma}}
\def\ba{\pmb{a}}
\def\bb{\pmb{b}}
\def\bd{\pmb{d}}

\def\bb{\pmb{b}}

\def\bx{\pmb{x}}
\def\by{\pmb{y}}

\def\bzero{\pmb{0}}

\def\bImat{{\rm {\bf I}}}

\begin{document}

\begin{center}

\begin{Large}

{\bf A LEVEL-SET HIT-AND-RUN SAMPLER FOR QUASI-CONCAVE DISTRIBUTIONS}

\end{Large}

\vspace{2cm}

\begin{large}
Dean Foster and Shane T. Jensen

\vspace{0.25cm}

Department of Statistics \\ The Wharton School\\ University of Pennsylvania \\
{\tt{stjensen@wharton.upenn.edu}}

\end{large}

\vspace{1cm}

\end{center}

\begin{abstract}
We develop a new sampling strategy that uses the hit-and-run algorithm within level sets of the target density.   Our method can be applied to any quasi-concave density, which covers a broad class of models.   Our sampler performs well in high-dimensional settings, which we illustrate with a comparison to Gibbs sampling on a spike-and-slab mixture model.  We also extend our method to exponentially-tilted quasi-concave densities, which arise often in Bayesian models consisting of a log-concave likelihood and quasi-concave prior density.  Within this class of models, our method is effective at sampling from posterior distributions with high dependence between parameters, which we illustrate with a simple multivariate normal example.  We also implement our level-set sampler on a Cauchy-normal model where we demonstrate the ability of our level set sampler to handle multi-modal posterior distributions.  
\end{abstract}

\vspace{1.5cm}

\begin{center}
\today

\end{center}

\newpage

\doublespacing

\noindent

\section{Introduction}

Complex statistical models are often estimated by sampling random variables from complicated distributions.  This strategy is especially prevalent within the Bayesian framework, where Markov Chain Monte Carlo (MCMC) simulation is used to estimate the  posterior distribution of a set of unknown parameters.   The most common MCMC technique is the Gibbs sampler \cite[]{GemGem84}, where small subsets (often one parameter at a time) are sampled from their conditional posterior distribution given the current values of all other parameters.  This component-wise strategy reduces the sampling from complicated multi-dimension distributions into sampling from a series of small (or one-) dimension distributions.   For sampling in low dimensions, various techniques such as the Metropolis-Hastings algorithm \cite[]{Has70} can be employed.  

However,  in high dimensional parameter spaces component-wise strategies can encounter problems such as high autocorrelation and the inability to move between local modes.   In high dimensions, one would ideally employ a strategy that sampled the high dimensional parameter vector directly, instead of one component at a time.  In this paper, we develop a sampling algorithm that provides direct samples from a (potentially high-dimensional) {\it quasi-concave} density function.  

Consider a high-dimensional density function $f(\bx)$ from which we would like to obtain samples $\bx$.  A density function $f$ is quasi-concave if:
$$C_a = \{ x \, | \, f(x) > a \}$$
is convex for all values of $a$.   Our procedure is based on the fact that any horizontal slice through a quasi-concave density $f$ will give a convex level set above that slice.   By slicing the quasi-concave density $f$ at a sequence of different heights, we divide the density $f$ into a sequence of convex level sets.   

We then use the hit-and-run algorithm to sample high-dimensional points $\bx$ within each of these convex level sets, while simultaneously estimating the volume of each convex level set.   As reviewed in \cite{Vem05}, recent work suggests that the hit-and-run algorithm is an efficient way to sample from high-dimension convex sets as long as the start is ``warm".  The hit-and-run algorithm is not as commonplace as other sampling methods (e.g. Metropolis-Hastings), though \cite{CheSch96} discuss using hit-and-run Monte Carlo to evaluate multi-dimensional integrals.

In Section~\ref{warmtheory}, we review several recent results for hit-and-run sampling in convex sets.  In Section~\ref{warmsampleroutline}, we outline our procedure for ensuring a warm start within each convex slice, thereby giving us an efficient way of sampling from the entire quasi-concave density.  In Section~\ref{empirical1}, we present an empirical comparison that suggests our ``level-set hit-and-run sampling" methodology is much more efficient than component-wise sampling schemes for high dimensional quasi-concave densities.    

We also extend our method to sample efficiently from {\it exponentially-tilted quasi-concave} densities in Section~\ref{exponentialtiltedwarmsampler}.   Exponentially-tilted quasi-concave densities are very common in Bayesian models: the combination of a quasi-concave prior density and a log-concave likelihood leads to an exponentially-tilted quasi-concave posterior density.   

In Section~\ref{empirical2a}, we implement our exponentially-tilted level-set sampler on a simple multivariate normal density.    In high dimensions, our method is more effective at obtaining posterior samples than a component-wise Gibbs sampling strategy.  

Finally, in Section~\ref{empirical2}, we illustrate the efficiency of our method on one such model, consisting of a Normal data likelihood and a Cauchy prior density.   It should be noted that this posterior density can be multi-modal, depending on the relative locations of the prior mode and observed data.    Our results in Section~\ref{empirical2} suggest that our extended method can accommodate posterior distributions which are both high-dimensional and multi-modal.

\section{Methodology and Theory}\label{methods}

\subsection{Quasi-Concave Densities and Sampling Convex Level Sets} \label{warmtheory}

Let $f()$ be a density in a high dimensional space, then $f$ is quasi-concave if:
$$C_a = \{ \bx \, | \, f(\bx) > a \}$$
is convex for all values of $a$.  In other words, the level set $C_a$ of a quasi-concave density $f$ is convex for any value $a$.  Let $\mathcal{Q}$ denote the set of all quasi-concave densities and $\mathcal{D}$ denote the set of all concave densities.  
All concave densities are quasi-concave, $\mathcal{D} \subset \mathcal{Q}$, but the converse is not true.    Quasi-concave densities are a very broad class that contains many common data models, including the normal, gamma, student's $t$, and uniform densities.   

Our proposed methodology is based on computing the volume of a convex level set $C$ by obtaining random samples $\bx$ from $C$.   Let us assume we are already at a point $\bx^0$ in $C_a$.   A simple algorithm for obtaining a random sample $\bx$ would be a {\it ball walk}: pick a uniform point $\bx$ in a ball of size $\delta$ around $\bx^0$, and move to $\bx$ only if $\bx$ is still in $C$.   

An alternative algorithm is {\it hit-and-run}, where a random direction $\bd$ is picked from current point $\bx^0$.   This direction will intersect the boundary of the convex set $C$ at some point $\bx^1$.   A new point $\bx$ is sampled uniformly along the line segment defined by direction $\bd$ and end points $\bx^0$ and $\bx^1$, and is thus guaranteed to remain in $C$.  

\cite{Lov99} showed that the hit-and-run algorithm mixes rapidly from a {\it warm start} in a convex body.  The warm start criterion is designed to ensure that our starting point is not stuck in some isolated corner of the convex body.   

\cite{Vem05} suggests that a random point from convex set $C^\prime \in C$ provides a warm start for convex set $C$ as long as ${\rm volume}(C^\prime) / {\rm volume} (C) \geq 0.5$.    \cite{Vem05} also presents several results that suggest the hit-and-run algorithm mixes more rapidly than the ball walk algorithm.

\subsection{LSHR1: Level-Set Hit-and-run Sampler for Quasi-concave Densities}\label{warmsampleroutline}

We harness the results of the previous section in our sampling algorithm for an arbitrary quasi-concave probability density.   Briefly,  our algorithm proceeds in stages where we sample from increasingly larger level sets of the quasi-concave density, while ensuring that each level set provides a warm start for the next level set.   After the entire density has been covered in this fashion, our samples from each level set are weighted appropriately to provide a full sample from the quasi-concave probability density.   We will see in later sections that our procedure is far more efficient than standard MCMC methods when the desired quasi-concave density has high dimension.  We outline our algorithm in more detail below, and also include pseudo-code in Appendix~\ref{warmsamplerdetails}.

Our algorithm starts by taking a small level set centered around the maximum value $\bx_{\max}$ of the quasi-concave density $f(\cdot)$.   This initial slice is defined as the convex shape $C_1$ of the probability density $f(\cdot)$ above an initial density threshold $t_1$.  In other words, a particular point $\bx$ is a valid member of slice $C_1$ iff $f(\bx) \geq t_1$.   

Starting from $\bx_{\max}$, we run a hit-and-run sampler within this initial level set for $m$ iterations.   In our supplementary materials, we explore appropriate choices for the number of hit-and-run iterations $m$.  

Each iteration of the hit-and-run sampler consists of first picking a random direction\footnote{Note that we can increase the efficiency of the hit-and-run sampler for highly non-spherical convex shapes by scaling this random direction by the estimated covariance matrix of our samples.  See Appendix~\ref{warmsamplerdetails} for details.} $\bd$ from current point $\bx$.  This direction $\bd$ and current point $\bx$ define a line segment with endpoints at the edge of the convex level set $C_1$.   A new point $\bx^\prime$ is sampled uniformly along this line segment, which ensures that $\bx^\prime$ remains in $C_1$.   We use $\{ \bx \}_1$ to denote the set of all points sampled from $C_1$ using this hit-and-run method.  

After sampling $m$ times from convex level set $C_1$, we need to pick a new threshold $t_2 <  t_1$ that specifies a new level set of the quasi-concave probability density $f(\cdot)$.   However, in order for our sampler to stay ``warm" as defined in Section~\ref{warmtheory}, we need $t_2$ to define a convex shape $C_2$ with volume $V_2$ that is less than twice the volume of our initial volume $V_1$.   

We define the volume ratio of two level sets as $R_{1:2} = V_1 / V_2$, so we need $R_{1:2} \geq 0.5$ for the two level sets $C_1$ and $C_2$.    In addition, we want a threshold $t_2$ that is not too close to $t_1$ so that we are not using an excessive number of level sets.

We meet these criteria by first proposing a new threshold $t_{\rm prop}$ and then, starting from current point $\bx$, running $m$ iterations of the hit-and-run sampler within the new convex level set $C_{\rm prop}$ defined by $f(\bx) > t_{\rm prop}$.  We focus on the ratio of volumes $R_{1:{\rm prop}} = V_1 / V_{\rm prop}$ which we estimate with $\hat{R}_{\rm prop}$ equal to the proportion of these sampled points from convex level set $C_{\rm prop}$ that were also contained within the previous level set $C_1$.   

We only accept the proposed threshold $t_{\rm prop}$ if $0.55 \leq \hat{R}_{\rm prop} \leq 0.8$.   The lower bound fulfills the criterion for the new level to be ``warm" and the {\it ad hoc} upper bound fulfills our desire for efficiency: we do not want the new level set to not be too close in volume to the previous level set.   

If the proposed threshold is not accepted, we propose a new threshold $t_{\rm prop}$ in an adaptive fashion (details in Appendix~\ref{warmsamplerdetails}) and repeat the same steps for testing this new threshold.   If the proposed threshold is accepted, we re-define $t_{\rm prop} \equiv t_2$ and then the convex level set $C_{\rm prop} \equiv C_2$ becomes our current level set.  The collection of sampled points $\{ \bx \}_2$ from $C_2$ are retained, as is our estimated ratio of slice volumes $\hat{R}_{1:2}$.   

In this same way, we continue to add level sets $C_k$ defined by threshold $t_k$ based on comparison to the previous level sets $C_{k-1}$.   The level set algorithm terminates when the accepted threshold $t_k$ is lower than some pre-specified lower bound $K$ on the probability density $f(\cdot)$.    

The result of this entire procedure is $n$ level sets, represented by a vector of thresholds  $(t_1,\ldots t_n)$, a vector of estimated ratios of volumes $(\hat{R}_{1:2},\ldots,\hat{R}_{n-1:n})$, and the level set collections: a set of $m$ sampled points from each level set $( \{ \bx \}_1, \ldots, \{ \bx \}_n)$.  

In order to obtain $L$ essentially independent\footnote{Although there is technically some dependence between successive sampled points from the hit-and-run sampler, our scheme of sub-sampling randomly from level set collections $( \{ \bx \}_1,  \ldots, \{ \bx \}_{n})$ essentially removes this dependence} samples from the pseudo-convex density, we sub-sample points from the level set collections $( \{ \bx \}_1,  \ldots, \{ \bx \}_{n})$, with each level set $i$ represented proportional to its probability $p_i$.  

The probability of each level set is the product of its density height and volume,
$$p_i = (t_{i-1}-t_{i}) \times V_i$$
We have our volume ratios $\hat{R}_{i:i+1}$ as estimates of $V_i / V_{i+1}$, which allows us to estimate $p_i$ by first calculating 
\begin{eqnarray}
\hat{q}_i \,\, = \,\, (t_{i-1}-t_{i}) \times \prod\limits_{j=i}^{n} \hat{R}_{j:j+1}    \qquad \qquad i = 1,\ldots,n \label{probcalc}
\end{eqnarray}
where $t_{0}$ is the maximum of $f(\cdot)$ and $\hat{R}_{n:n+1} = 1$, and then $\hat{p_i} = \hat{q}_i / \sum \hat{q}_j$.   

We demonstrate our LSHR1 algorithm on a spike-and-slab density in Section~\ref{empirical1}.

\subsection{Exponentially-Tilted Quasi-concave Densities}\label{exponentialtilteddensities}

A density $g()$ is a log-concave density function if $\log g()$ is a concave density function.  Let $\mathcal{L}$ denote the set of all log-concave density functions.   All log-concave density functions are also concave density functions ($\mathcal{L} \subset \mathcal{D}$), and thus all log-concave density functions are also quasi-concave density functions ($\mathcal{L} \subset \mathcal{Q}$).    

\cite{bagber05} gives an excellent review of log-concave densities.  The normal density is log-concave whereas the student's $t$ and Cauchy density are not.   The gamma and beta densities are log-concave only under certain parameter settings, e.g. both the uniform and exponential densities are log-concave.  The beta($a,b$) density with other parameter settings (e.g. $a < 1, b < 1$) can be neither log-concave nor quasi-concave.  

Now, let $T$ denote the set of {\it exponentially-tilted} quasi-concave density functions.  A density function $h$ is an exponentially-tilted quasi-concave density function if there exists a quasi-concave density $f \in \mathcal{Q}$ such that $f(\bx)/h(\bx) = \exp(\bbeta^\prime \bx)$.  Exponentially-tilted quasi-concave densities are a generalization of quasi-concave densities, so $\mathcal{Q} \subset \mathcal{T}$.  

These three classes of functions (log-concave, quasi-concave, and exponentially-tilted quasi-concave) are linked by the following important relationship:  if $X \sim f$ where $f \in \mathcal{Q}$ and $Y | X \sim g$ where $g \in \mathcal{L}$, then $X | Y \sim h$ where $h \in \mathcal{T}$.   

The consequences of this relationship is apparent for the Bayesian approach.  The combination of a quasi-concave prior density for parameters $\btheta$ and a log-concave likelihood for data $\by | \btheta$ will produce an exponentially-tilted quasi-concave posterior density $\btheta | \by$.   

We now extend our level-set hit-and-run sampling methodology to exponentially-tilted quasi-concave densities, which makes our procedure applicable to a large class of Bayesian models consisting of quasi-concave priors and log-concave likelihoods.  

As mentioned above, examples of log-concave likelihoods for data $\by | \btheta$ include the normal density, exponential density and uniform density.  Quasi-concave priors for parameters $\btheta$ are an even broader class of densities, including the normal density, the $t$ density, the Cauchy density and the gamma density.  

In Section~\ref{empirical2a}, we examine a simple multivariate normal density, which is log-concave.  Then,  in Section~\ref{empirical2},  we examine an exponentially-tilted quasi-concave posterior density resulting from the combination of a multivariate normal density with a Cauchy prior distribution. 
 
 \subsection{LSHR2: Level-Set Hit-and-Run Sampler for Exponentially-Tilted Quasi-concave Densities}\label{exponentialtiltedwarmsampler}

In Section~\ref{warmsampleroutline}, we presented our level set hit-and-run algorithm for sampling $\bx$ from quasiconcave density $f(\bx)$.   We now extend our algorithm to sample from an exponentially-tilted quasi-concave density $h(\bx)$.   

As mentioned in Section~\ref{exponentialtilteddensities}, exponentially-tilted quasi-concave densities commonly arise as posterior distributions in Bayesian models.   In keeping with the usual notation for Bayesian models, we replace our previous variables $\bx$ with parameters $\btheta$.  These parameters $\btheta$ have an exponentially-tilted quasi-concave posterior density $h(\btheta | \by)$  arising from a log-concave likelihood $g(\by | \btheta)$ and quasi-concave prior density $f(\btheta)$.  

Our LSHR2 algorithm starts just as before, by taking a small level set centered around the maximum value $\btheta_{\max}$ of the quasi-concave prior density $f(\cdot)$.  This initial level set is defined, in part, as the convex shape of the probability density $f(\cdot)$ above an initial density threshold $t_1$.  

However, we now augment our sampling space with an extra variable $p$ where $p < g(\by | \log \btheta)$.  Letting $\btheta^\star = (\btheta,p)$, our convex shape is now 
$$ D_1 = \{ \btheta^\star: f(\btheta) \geq t_1 \, , \, p < \log g(\by | \btheta) \} $$

Within this new convex shape, we run an exponentially-weighted version of the hit-and-run algorithm.  Specifically, a random ($d+1$)-dimensional direction $\bd$ is sampled\footnote{Again we can increase the efficiency of the hit-and-run sampler for highly non-spherical convex shapes by scaling this random direction by the estimated covariance matrix of our samples.  See Appendix~\ref{expwarmsamplerdetails} for details.} which, along with the current $\btheta^\star$, defines a line segment with endpoints at the boundaries of $D_1$.   Instead of sampling uniformly along this line segment, we sample a new point $(\btheta^\star)^\prime$ from points on this line segment weighted by $\exp(p)$.   

The remainder of the LSHR2 algorithm proceeds in the same fashion as our LSHR1 algorithm:  we construct a schedule of decreasing thresholds $t_1,t_2,\ldots$ and corresponding level sets $D_1, D_2, \ldots$ such that each level set $D_k$ is a warm start for the next level set $D_{k+1}$.   

Within each of these steps, the exponentially-weighted hit-and-run algorithm is used to sample values $\btheta^\star$ within the current level set.  As before, $D_k$ is a warm start for the next level set $D_{k+1}$ if the ratio of volumes\footnote{We continue to use the term ``volume" even though these convex shapes are a combination of $d$-dimensional $\theta$ and the extra log-density $p$ dimension.} $\hat{R}_{k:k+1} = V_k / V_{k+1}$ that lies within the range $0.55 \leq \hat{R}_{k:k+1}  \leq 0.8$  The algorithm terminates when our decreasing thresholds $t_k$ achieve a pre-specified lower bound $K$.  

Our procedure results in $n$ level sets $(D_1,\ldots,D_n)$, represented by a vector of thresholds  $(t_1,\ldots t_n)$, a vector of estimated ratios of volumes $(\hat{R}_{1:2},\ldots,\hat{R}_{n-1:n})$, and the level set collections: a set of $m$ sampled points from each level set: 
$$( \{ \btheta^\star \}_1, \ldots, \{ \btheta^\star \}_n)$$    
Finally, we obtain $L$ samples $(\btheta^\star_1,\ldots,\btheta^\star_L)$ by sub-sampling points from the level set collections $( \{ \btheta^\star \}_1, \ldots, \{ \btheta^\star \}_n)$, with each level set $i$ represented proportional to its probability $p_i$. The probability of each level set is still calculated as in (\ref{probcalc}).  

By simply ignoring the $(d+1)$-th sampled dimension $p$, we have samples $(\btheta_1,\ldots,\btheta_L)$ from the exponentially-tilted pseudo-convex posterior density $h(\btheta | \by)$.     Details of our LSHR2 algorithm are given in Appendix~\ref{expwarmsamplerdetails}.    We demonstrate our LSHR2 algorithm on a multivariate normal density in Section~\ref{empirical2a} and a Cauchy-normal model in Section~\ref{empirical2}.  

\subsection{Component-wise Alternatives:  Gibbs and Metropolis-Hastings}

Although there have been many recent advances in MCMC methods, we focus on our comparisons on the Gibbs sampler as it remains the most popular method for obtaining samples from high-dimensional distributions.   The Gibbs sampler \cite[]{GemGem84} is a Markov Chain Monte Carlo approach to sampling from a multi-dimensional density $f(\bx)$.   

The basic Gibbs strategy is to sample a single component $x_i$ from its univariate conditional distribution $f(x_i | \bx_{-i})$ given all other components $\bx_{-i}$.   Successively iterating this sampling strategy through each component $i = 1,\ldots,n$ will give a full sample $\bx$.  These sampled $\bx$'s will converge in distribution to the target distribution $f(\bx)$.   

The benefit of a Gibbs approach is that sampling from a high-dimensional density $f(\bx)$ is reduced to sampling from a series of one-dimensional densities $f(x_i | \bx_{-i})$.   The Gibbs sampler is most attractive when these one-dimensional conditional distributions each have a standard form but even if this is not the case, the Metropolis-Hastings algorithm \cite[]{Has70} can be employed for sampling from non-standard distributions.   

The Gibbs sampler is not always strictly component-wise: for some models, blocks of variables can be sampled in a single step from a standard distribution.   That said, a Gibbs strategy for the sampling of a high-dimensional $\bx$ will typically involve sampling a series of small subsets of $\bx$. 

The component-wise approach of the Gibbs sampler has both strengths and weaknesses.  The relative simplicity of sampling from conditional distributions is a clear strength.    However, a potential weakness is that  high correlation can lead to extremely slow convergence to the target distribution.  Our own approach is motivated by the belief that it is advantageous to sample the entire set of variables simultaneously when possible.

\section{Empirical Study 1: Spike-and-slab Density}\label{empirical1}

\begin{figure}[ht!]
\caption{Spike-and-slab density: gray lines indicate density of the two components, black line is the mixture density}
\label{spikeslab}
\begin{center}
\includegraphics[width=5in]{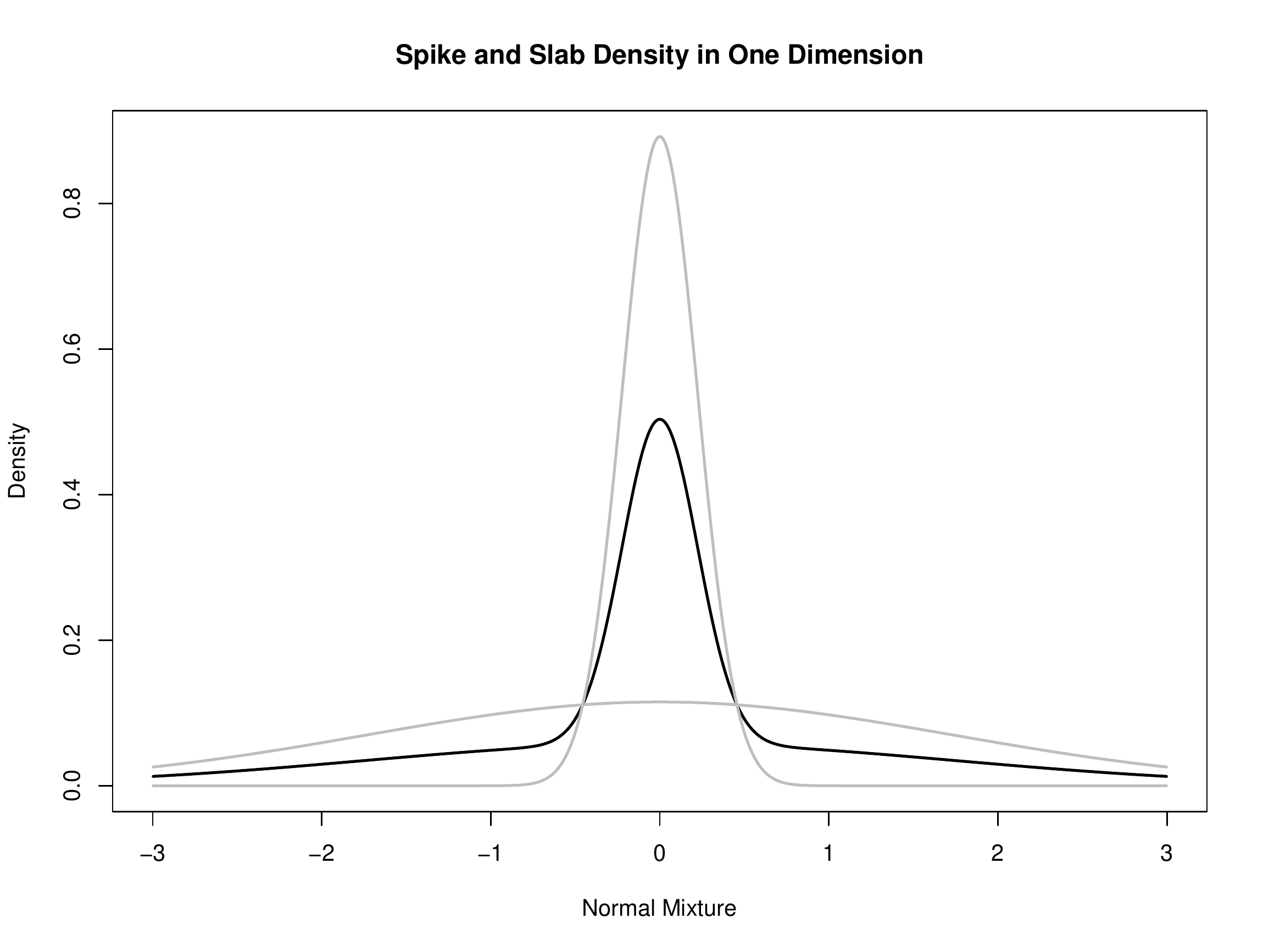}
\end{center}
\end{figure}

We consider the problem of sampling from a multivariate density which consists of a 50-50 mixture of two normal distributions, both centered at zero but different variances.  Specifically, the first component has variance $\bSigma_0$ with off-diagonal elements equal to zero and diagonal elements equal to 0.05, whereas the second component has variance $\bSigma_0$ with off-diagonal elements equal to zero and diagonal elements equal to 3.  Figure~\ref{spikeslab} gives this spike-and-slab density in a single dimension.  
The spike-and-slab density is commonly used as a mixture model for important versus spurious predictors in variable selection models \cite[]{GeoMcC97}.    

This density is quasi-concave and thus amenable to our proposed level-set hit-and-run sampling methodology.  Sampling from this density in a single dimension can be easily accomplished by grid sampling or Gibbs sampling using data augmentation.  However, we will see that these simpler strategies will not perform adequately in higher dimensions whereas our LSHR1 sampling strategy continues to show good performance in high dimensions.   

\subsection{Gibbs Sampling for Spike and Slab}

The usual Gibbs sampling approach to obtaining a sample $\bx$ from a mixture density is to augment the variable space with an indicator variable $I$ where $I = 1$ indicates the current mixture component from which $x$ is drawn.  The algorithm iterates between 
\begin{enumerate}
\item Sample $x$ from mixture component dictated by $I$, i.e. 
\[ \bx \sim \left\{ \begin{array}
           {r@{\qquad {\rm if} \quad}l@{ \quad}l}
          {\rm N} (\bzero, \bSigma_1) & I = 1 &\\
         {\rm N} (\bzero, \bSigma_0) & I = 0 & \label{mixture} \\ 
           \end{array} \right. \]  
\item Sample $I$ with probability based on $x$:  $P(I = 1) =  \frac{\phi (\bx, \bzero, \bSigma_1)}{\phi (\bx, \bzero, \bSigma_1) + \phi (\bx, \bzero, \bSigma_1)}$
\end{enumerate}
where $\phi (\bx, \bmu, \bSigma)$ is the density of a multivariate normal with mean $\mu$ and variance $\bSigma$ evaluated at $\bx$.   

\begin{figure}[ht!]
\caption{Empirical Mixing Proportion after each iteration is calculated as the proportion of samples up to that iteration that are from the first component (the spike).  Black lines indicate the empirical mixing proportion for the Gibbs sampler, gray lines indicate the true mixing proportion of 0.5}
\label{gibbsmixing}

\vspace{-0.5cm}

\begin{center}
\includegraphics[width=6in]{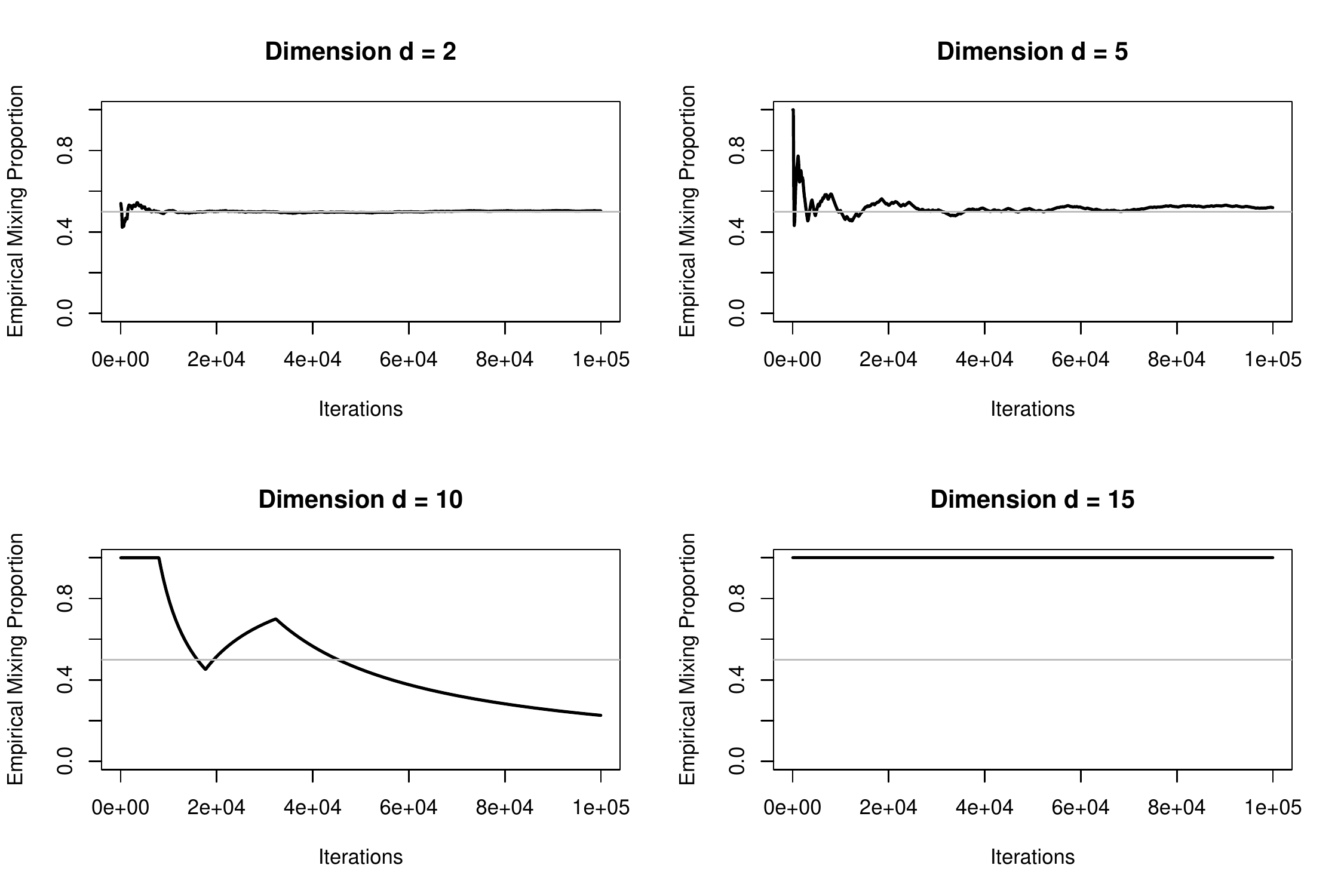}
\end{center}
\end{figure}

This algorithm mixes well when $\bx$ has a small number of dimensions ($d < 10$), but in higher dimensions, it is difficult for the algorithm to move between the two components.   We see this behavior in Figure~\ref{gibbsmixing}, where we evaluate results from running the Gibbs sampler for 100000 iterations on the spike-and-slab density with different dimensions ($d=2,5,10,15$).   Specifically, we plot the proportion of the Gibbs samples taken from the first component (the spike) for $\bx$ of different dimensions.   

\begin{figure}[ht!]
\caption{Number of component switches is the number of times during each Gibbs sampler run (out of 100000 iterations) that the algorithm moved from one mixture component to the other.}
\label{gibbsswitches}

\vspace{-1cm}

\begin{center}
\includegraphics[width=5in]{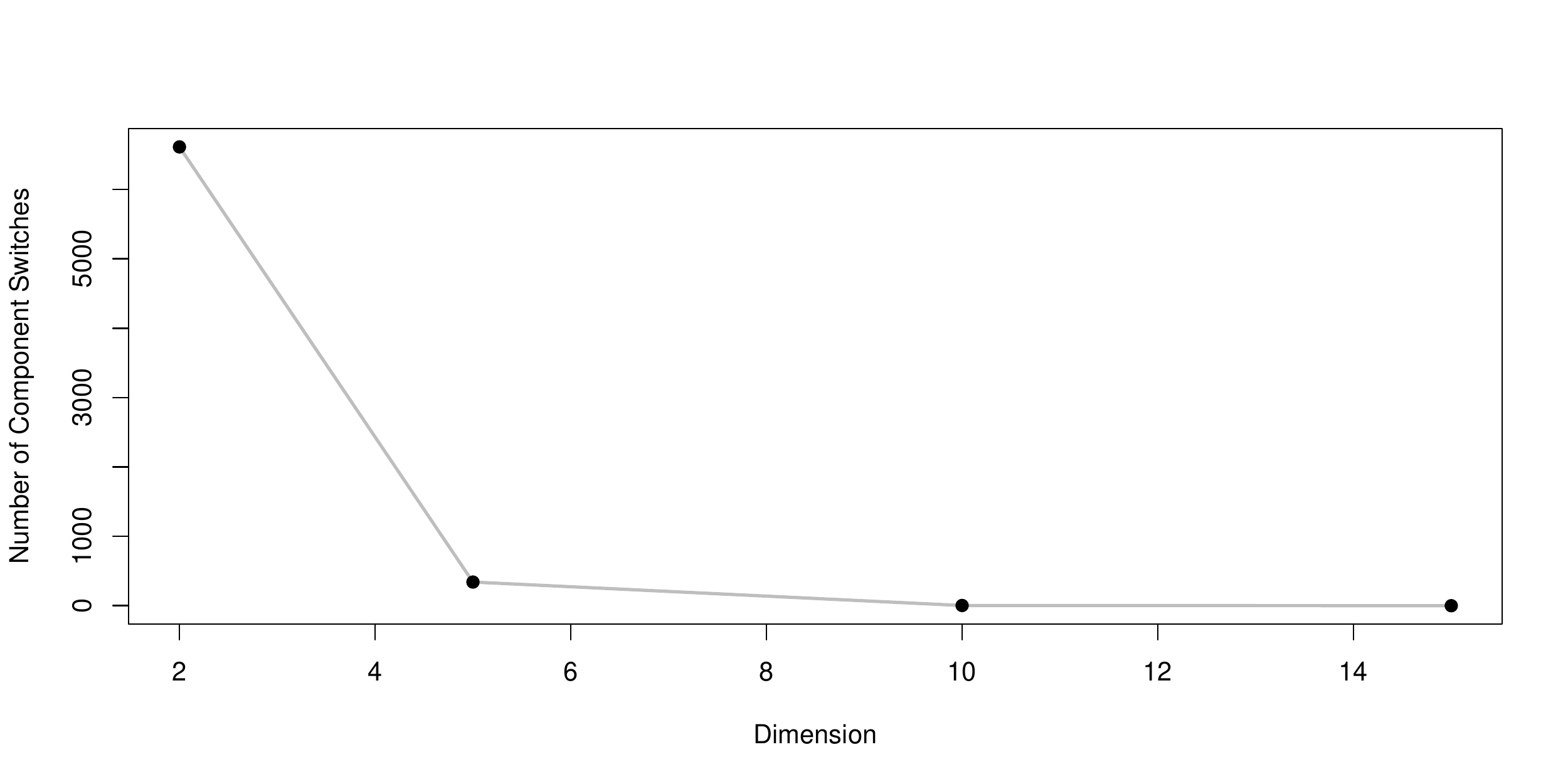}
\end{center}
\end{figure}

The mixing of the sampler for lower dimensions ($d = 2$ and $d = 5$) is reasonable, but we can see that for higher dimensions the sampler is extremely sticky and does not mix well at all.   In the case of $d=10$, the sampler only makes a couple movies between the spike component and the slab component  during the run.   In the case of $d=15$, the sampler does not move into the second component at all during the entire 100000 sample run.  We see this behavior more explicitly in Figure~\ref{gibbsswitches} where we plot the number of switches between components from the Gibbs sampling run for each dimension. Even higher dimensions ($d=20,d=25$) were similar to the $d=15$ case.  

In high dimensions, we can estimate how often the Gibbs sampler will
switch from one domain to the other.  When the sampler is in one of
the components, the variable $x$ will mostly have a norm of $||{\bf
x}||_2^2 \approx
d \sigma^2$ where $\sigma = \sigma_0 = .05$ for component zero and
$\sigma = \sigma_1 = 3.0$
for component one.  For a variable currently classified into the zero component, the probability of a switch is  
\begin{eqnarray*}
P\left(I = 1 \, | \, ||{\bf x}||_2^2  \approx
d \sigma_0^2\right)  \approx
\left(\frac{\sigma_0\sqrt{e}}{\sigma_1}\right)^d
\end{eqnarray*}
where the approximation holds if $\sigma_0 \ll \sigma_1$.  Details of this calculation are given in Appendix~\ref{gibbsswitch}.  This result suggests that the expected number of iterations until a switch is about $(\sigma_1/\sigma_0\sqrt{e})^d =
36.4^d$.  This approximate switching time will be more accurate for large
$d$.

\subsection{Level-set Hit-and-run Sampling for Spike and Slab}

Our LSHR1 sampling methodology was implemented on the same spike-and-slab distribution.  Our sampling algorithm does not depend on a data augmentation scheme that moves between the two components.  Rather, we start from the posterior mode and move outwards in convex level sets, while running a hit-and-run algorithm within each level set.   As outlined in Section~\ref{exponentialtiltedwarmsampler}, each convex level set is defined by a threshold on the density that is slowly decreased in order to assure that each level set has a warm start. 

\begin{figure}[ht!]
\caption{Number of thresholds needed for our level-set hit-and-run sampler on the spike-and-slab density as function of dimension.}
\label{spikeslabnumthres}

\vspace{-1cm}

\begin{center}
\includegraphics[width=3.75in]{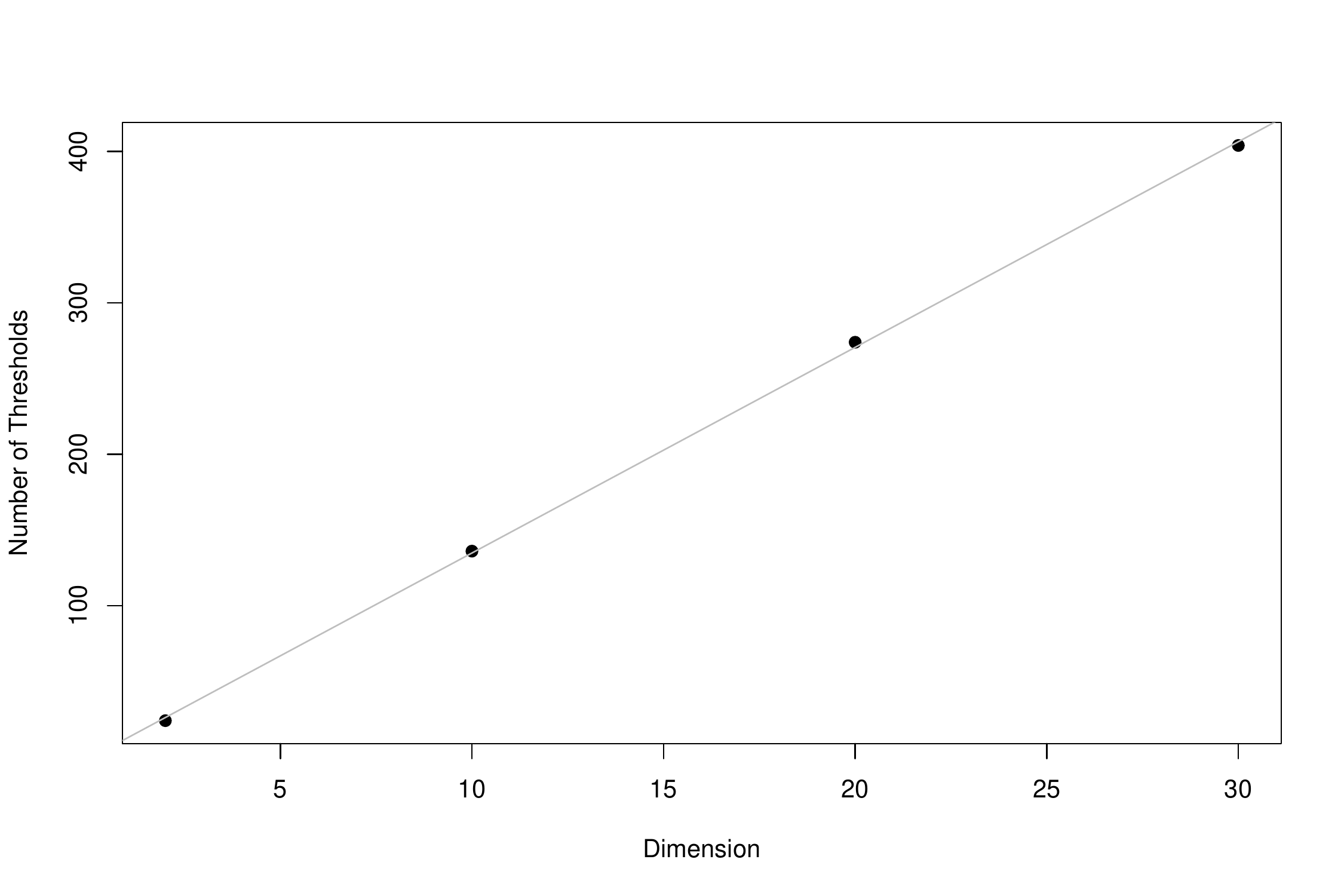}
\end{center}
\end{figure}

Figure~\ref{spikeslabnumthres} shows the number of thresholds needed to fully explore the spike-and-slab density.  Even for high dimensions ($d=10$ and $d=20$), we see that the spike-and-slab density is explored with a reasonable number of thresholds, and the linear trend suggests that our algorithm would scale well to even higher dimensions.  

\begin{figure}[ht!]
\caption{Comparing level-set hit-and-run sampler and Gibbs sampler in terms of quantile-quantile plot of estimated vs. true spike-and-slab distribution.  Top row is the low dimension ($d=2$) case, bottom row is the high dimension ($d=20$) case. }
\label{quantiles}

\vspace{-0.5cm}

\begin{center}
\includegraphics[width=5.5in]{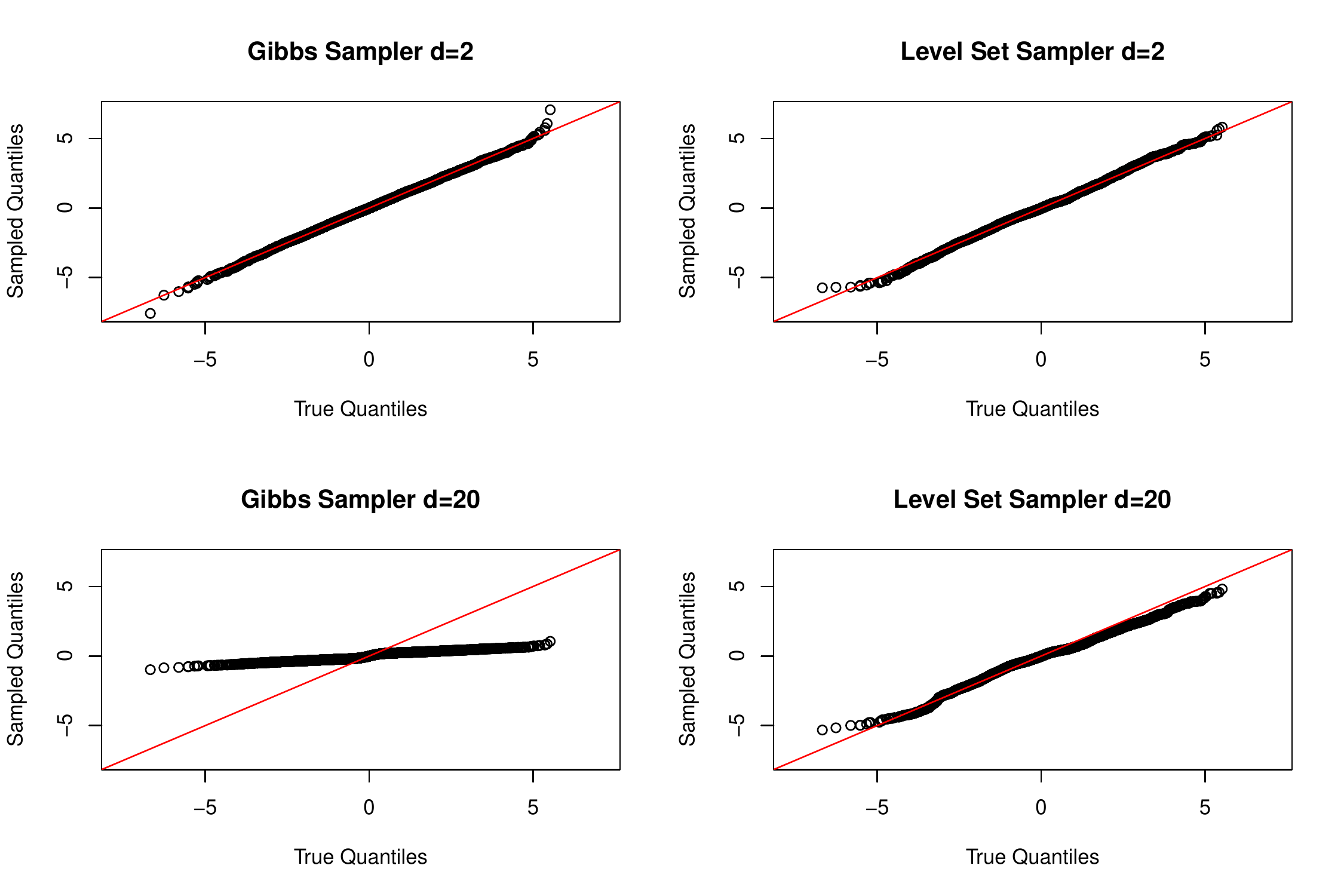}
\end{center}
\end{figure}

The most striking comparison of the superior performance of our LSHR1 sampling strategy comes from examining the sampled values in any particular dimension relative to the true quantiles of the spike-and-slab mixture density.  In Figure~\ref{quantiles}, we plot the true quantiles against the quantiles of the first dimension of sampled values from both the Gibbs sampler and the warm sampler.  

We see that for low dimensions ($d=2$), samples from both the Gibbs sampler and warm sampler match the correct spike-and-slab distribution.  However, for higher dimensions ($d=20$), the Gibbs sampler provides a grossly inaccurate picture of the true distribution, due to the fact that the Markov chain never escapes the spike component of the distribution.  In contrast, the samples from our level-set hit-and-run sampler still provide a good match to the true distribution in high dimensions.  We have checked even higher dimensions ($d=30$) and the level-set hit-and-run sampler still provides a good match to the true distribution.

\section{Empirical Study 2: Multivariate Normal}\label{empirical2a}

In our second empirical study, we compare our LSHR2 algorithm to the Gibbs sampler for obtaining samples from the multivariate normal density, 
\begin{eqnarray}
\btheta & \sim & {\rm Normal} (\bzero, \bSigma)   \label{normaldensity}
\end{eqnarray}
This is a superficial setting in the sense that it is easy to obtain samples $\btheta$ directly from a multivariate normal density.  However, this simple example is still illustrative of differences in performance between our LSHR2 level-set sampler and a component-wise Gibbs sampler.   

We considered several choices of dimension $d$ ($d= 2, 10, 20, 30$) and also several specifications of the data variance $\bSigma$.  Specifically, we assume all diagonal elements of $\bSigma$ are one, but we consider several different values for off-diagonal elements $\rho$ ($\rho = 0, 0.5, 0.9, 0.99$).  

We use our LSHR2 algorithm to obtain samples $\btheta$ by re-formulating (\ref{normaldensity}) as the exponentially-tilted quasi-concave posterior density that results from a log-concave multivariate normal likelihood of $\by = \bzero | \btheta, \bSigma$  and a bounded uniform prior density on $\btheta$ (i.e. $\theta_k \sim {\rm Unif}(-6,6)$ for $k=1,\ldots,d$).     In this simple situation where our quasi-concave prior density is uniform, we only have to use one level set in our LSHR2 algorithm.

We also use a component-wise Gibbs sampler to obtain samples of $\btheta$ by iteratively sampling from the conditional normal densities of $\theta_k | \btheta_{(-k)}$.    Both the Gibbs sampler and LSHR2 sampler were run for one million iterations.  In the case of the Gibbs sampler, the first 50000 iterations were discarded as burn-in.  

\begin{figure}[ht!]
\caption{Comparison of samples of $\btheta$ from multivariate normal density with $d=2$ and $\rho=0$.   Top row is samples from Gibbs sampler.  Bottow row is samples from LSHR2 algorithm.  Left column is the histograms of the  sampled values of $\theta_1$ (red line indicates true density).  Right column is the empirical autocorrelation functions of the sampled values of of $\theta_1$.  }
\label{normalexamplerho0}

\vspace{-0.5cm}

\begin{center}
\includegraphics[width=5.5in]{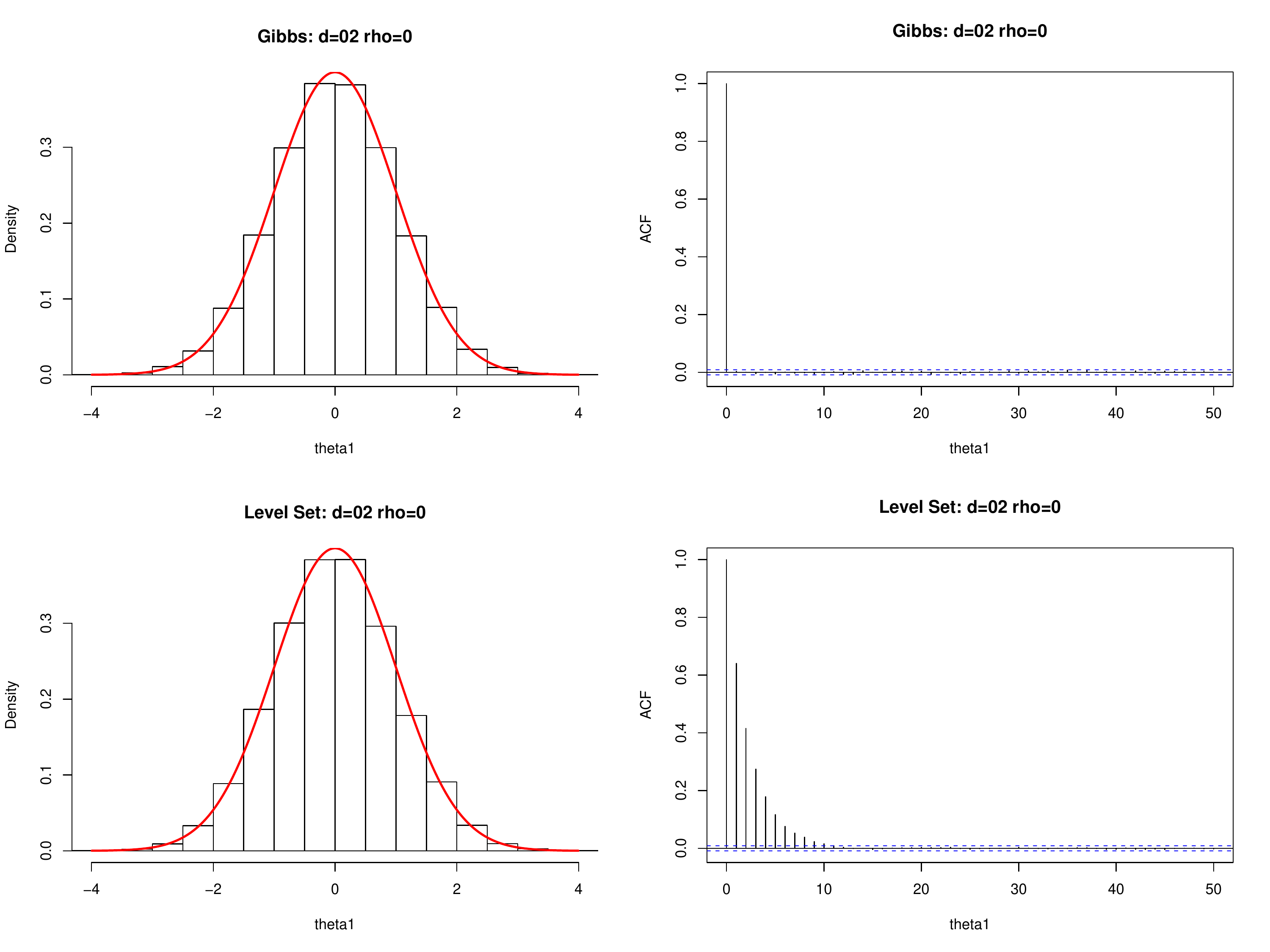}
\end{center}
\end{figure}

In Figure~\ref{normalexamplerho0}, we examine the performance of our LSHR2 algorithm versus the Gibbs sampler when $d=2$ and both dimensions are independent (i.e. $\rho = 0$).    Specifically, we give a histogram of the samples for the first dimension $\theta_1$ of $\btheta$ and the empirical autocorrelation function of the $\theta_1$ samples for both Gibbs and LSHR1 algorithms.   

\begin{figure}[ht!]
\caption{Comparison of samples of $\btheta$ from multivariate normal density with $d=2$ and $\rho=0.99$.   Top row is samples from Gibbs sampler.  Bottow row is samples from LSHR2 algorithm.  Left column is the histograms of the  sampled values of $\theta_1$ (red line indicates true density).  Right column is the empirical autocorrelation functions of the sampled values of of $\theta_1$.  }
\label{normalexamplerho99}

\vspace{-0.5cm}

\begin{center}
\includegraphics[width=5.5in]{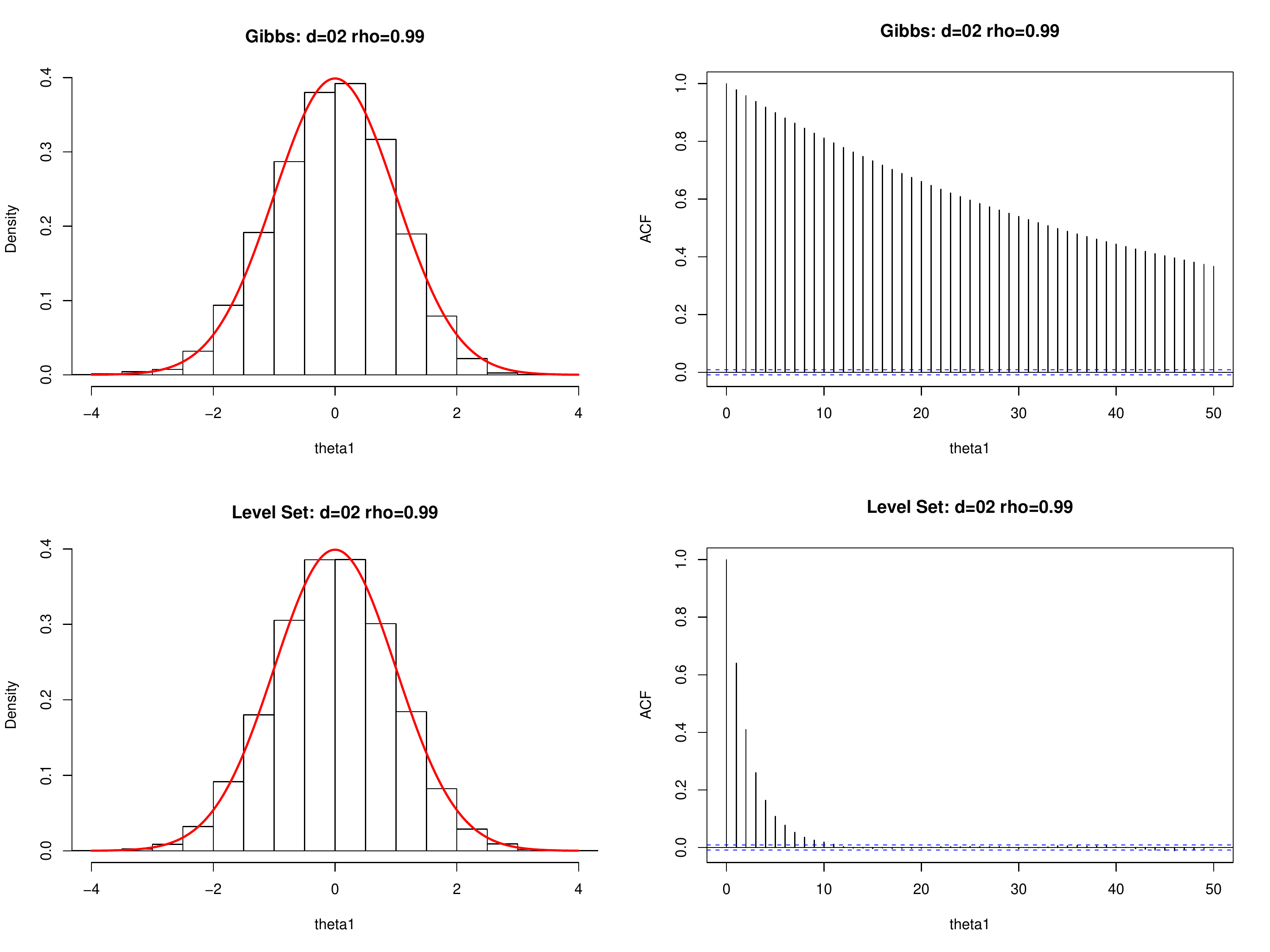}
\end{center}
\end{figure}

We see in Figure~\ref{normalexamplerho0} that the samples of $\theta_1$ from both algorithms are a close match to the true density.    We also see that the Gibbs sampler has less autocorrelation than our level-set procedure in this setting where the true correlation between the dimensions of $\btheta$ is zero.    Compare this observation with Figure~\ref{normalexamplerho99}, where we examine the performance of our LSHR2 algorithm versus the Gibbs sampler when $d=2$ but the correlation between the dimensions is high, i.e. $\rho = 0.99$.  

When the dimensions of $\btheta$ are highly correlated, we see in Figure~\ref{normalexamplerho0} that the performance of the Gibbs sampler is severely degraded, as evidenced by very high autocorrelation of sampled values.    In contrast, the autocorrelation of our LSHR2 algorithm is essentially unchanged between Figures~\ref{normalexamplerho0} and \ref{normalexamplerho99}, which indicates our methodology is robust to high dependence among the parameters $\btheta$.  

These same results were seen to an even greater extent when we compared the performance of the Gibbs sampler and LSHR2 algorithms in higher dimensions ($d= 10, 20, 30$).   Even in this artificial setting of a multivariate normal density with high correlation, we see that our level set methodology offers advantages over a component-wise Gibbs sampling strategy.  In the next section, we examine a more complicated model where our LSHR2 algorithm also out-performs Gibbs sampling.  

\section{Empirical Study 3: Cauchy-normal model}\label{empirical2}

In our third empirical comparison, we compare our LSHR2 algorithm to the Gibbs sampler on a Bayesian model consisting of a  multivariate normal likelihood and a Cauchy prior density, 
\begin{eqnarray}
\by | \btheta & \sim & {\rm Normal} (\btheta, \sigma^2 \bImat) \nonumber \\
\btheta & \sim & {\rm Cauchy} (\bzero, \bImat)   \label{cauchynormal}
\end{eqnarray}
where $\by$ and $\btheta$ are $d$ dimensions and $\sigma^2$ is fixed and known.   As mentioned in Section~\ref{exponentialtilteddensities}, this combination of a log-concave density $g(\by | \btheta)$ and a quasi-concave density for $f(\btheta)$ gives an {\it exponentially-tilted} quasi-concave posterior density $h(\btheta | \by)$.    

\begin{figure}[ht!]
\caption{Posterior density $h(\btheta | \by = 10)$ for Cauchy-normal model in (a) one and (b) two dimensions. }
\label{multimodal}

\vspace{-0.5cm}

\begin{center}
\includegraphics[width=6in]{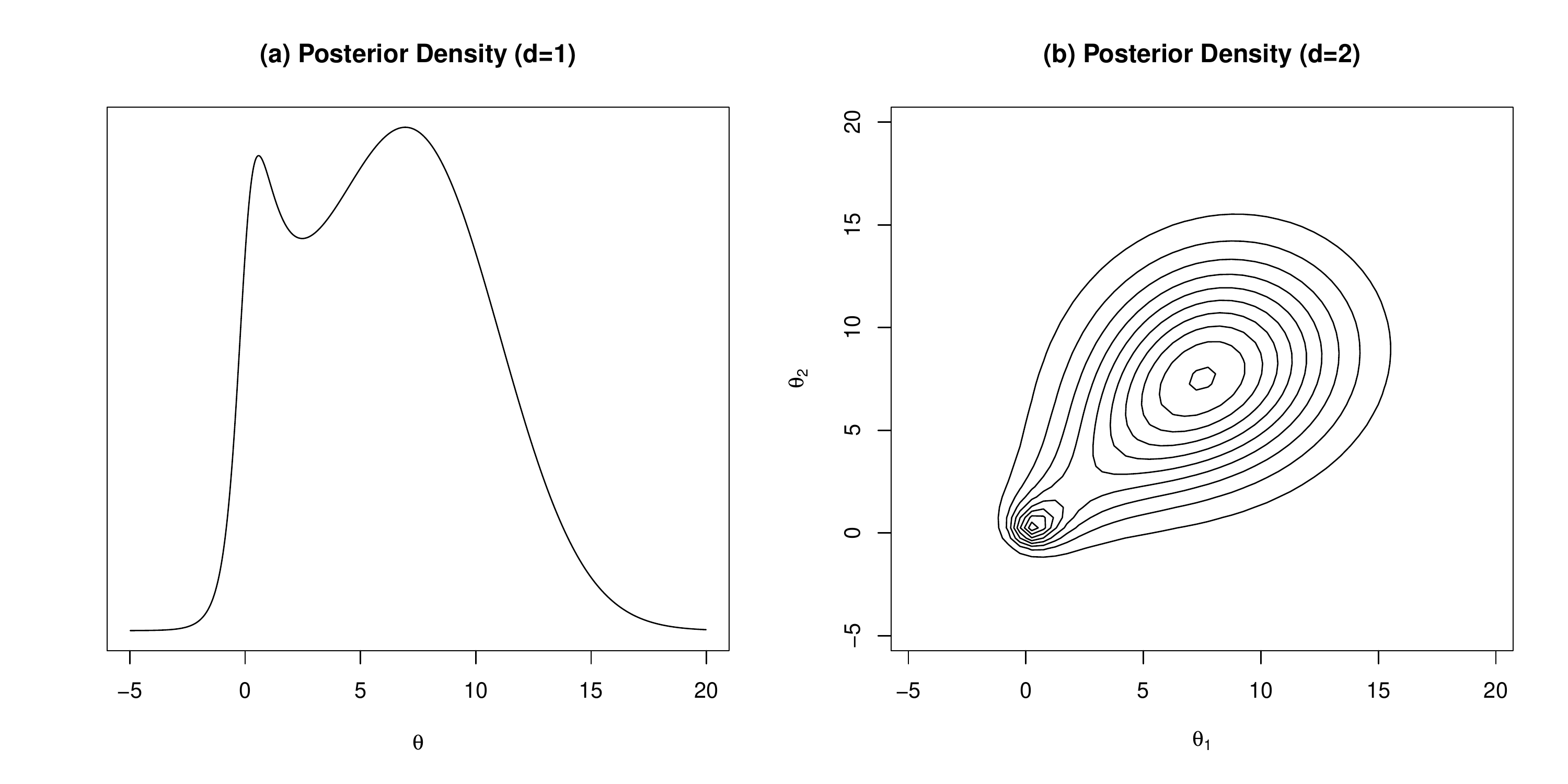}
\end{center}
\end{figure}

For some combinations of $\sigma^2$ and observed $\by$ values, the posterior density $h(\btheta | \by)$ is multi-modal.  Figure~\ref{multimodal} gives examples of this multi-modality in one and two dimensions.   The one-dimensional $h(\btheta | \by)$ in Figure~\ref{multimodal}a has $y = 10$ and $\sigma^2 = 10.84$, whereas the two-dimensional $h(\btheta | \by)$ in Figure~\ref{multimodal}b has $y = (10,10)$ and $\sigma^2 = 12.57$.   In higher dimensions $d$, we used $y = (10,\ldots,10)$ and $\sigma^2 = d\cdot10^2 / ((d+1)\log(1+d\cdot10^2))$, which is a formula derived from equating the density value at the prior mode to the density at the likelihood mode.   

\vspace{-0.75cm}

In higher dimensions ($d \geq 3$), it is difficult to evaluate (or sample from) the true posterior density $h(\btheta | \by)$ which we need in order to have a gold standard for comparison between the Gibbs sampler and LSHR2 algoirithm.   Fortunately, for this simple model, there is a rotation procedure that we detail in our supplementary materials that allows us to accurately estimate the true posterior density $h(\btheta | \by)$.  

\subsection{Gibbs Sampling for Cauchy Normal Model}

\begin{figure}[ht!]
\caption{Posterior samples of the first dimension $\theta_1$ of $\btheta$ from Gibbs sampler for Cauchy-normal model in different dimensions $d$. Red curve in each plot represents the true posterior density.}
\label{cauchynormalgibbshists}

\vspace{-0.5cm}

\begin{center}
\includegraphics[width=6in]{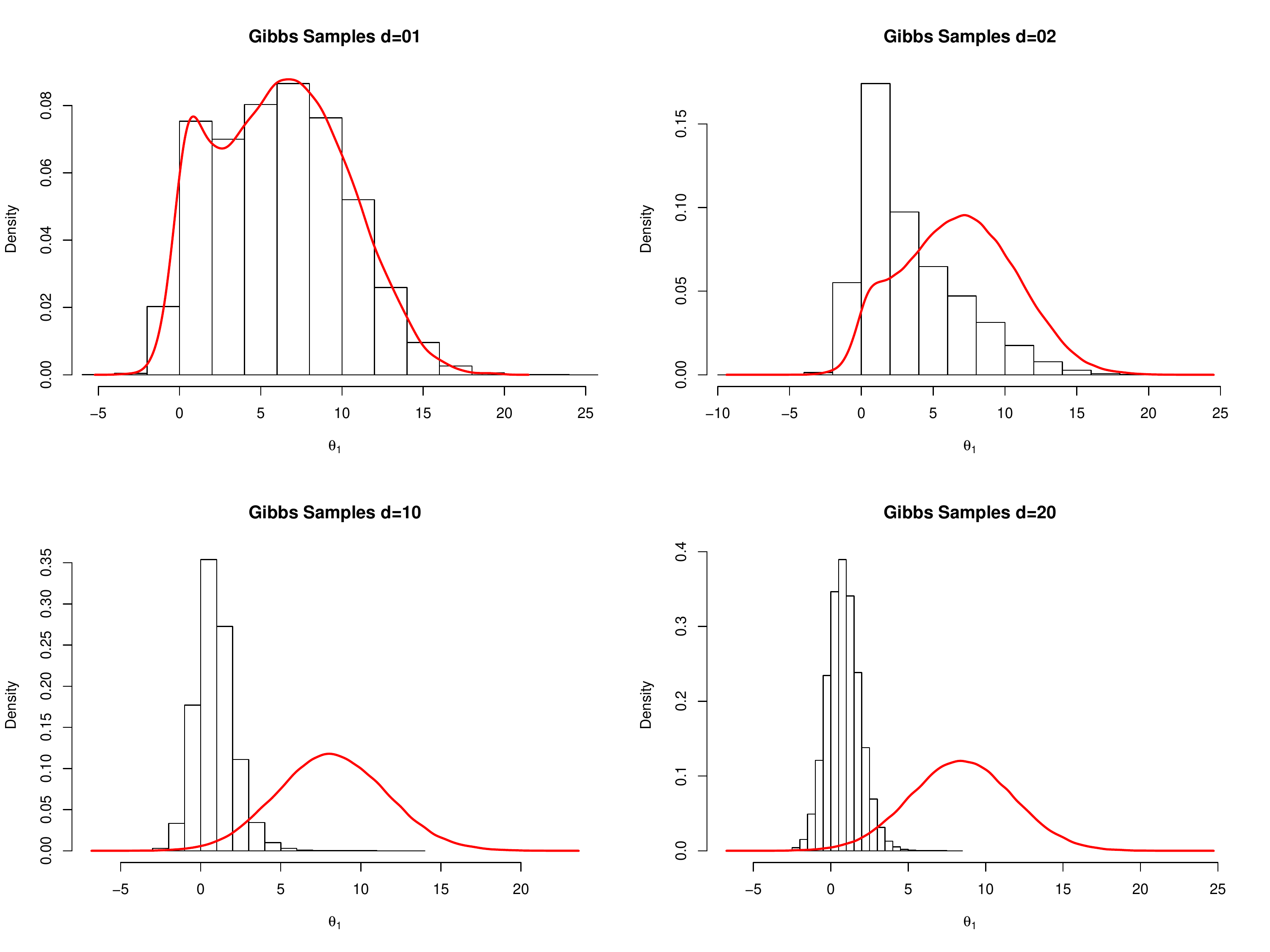}
\end{center}
\end{figure}

The Cauchy-normal model (\ref{cauchynormal}) can be estimated via Gibbs sampling by augmenting the parameter space with an extra scale parameter $\tau^2$, 
\begin{eqnarray}
\by | \btheta & \sim & {\rm Normal} (\btheta, \sigma^2 \bImat) \nonumber \\
\btheta | \tau^2 & \sim & {\rm Normal} (\bzero, \tau^2 \bImat)   \nonumber \\
(\tau^2)^{-1} & \sim & {\rm Gamma} \left(\frac{1}{2}, \frac{1}{2}\right) \label{cauchynormalgibbs}
\end{eqnarray}
Marginalizing over $\tau^2$ gives us the same Cauchy($\bzero, \bImat$) prior for $\btheta$.  The Gibbs sampler iterates between sampling from the conditional distribution of $\btheta | \tau^2, \by$: 
\begin{eqnarray}
\btheta | \tau^2, \by & \sim & {\rm Normal} \left(\frac{\frac{1}{\sigma^2}\by}{\frac{1}{\sigma^2} + \frac{1}{\tau^2}}  \, , \, \frac{1}{\frac{1}{\sigma^2} + \frac{1}{\tau^2}} \bImat \right) 
 \end{eqnarray}
and the conditional distribution of $\tau^2 | \btheta, \by$: 
\begin{eqnarray}
(\tau^2)^{-1} | \btheta, \by & \sim & {\rm Gamma} \left( \frac{d+1}{2} \, , \, \frac{d + \btheta^{\rm T} \btheta }{2}\right) 
\end{eqnarray}
For several different dimensions $d$, we ran this Gibbs sampler for one million iterations, with the first 500000 iterations discarded at burn-in.   In Figure~\ref{cauchynormalgibbshists}, we compare the posterior samples for the first dimension $\theta_1$ from our Gibbs sampler to the true posterior distribution for $d = 1, 2, 10$ and 20.   

We see that the Gibbs sampler performs poorly at exploring the full posterior space in dimensions greater than one.   Even in two dimensions, the Gibbs sampler struggles to fully explore both modes of the posterior distribution and this problem is exacerbated in dimensions higher than two.   

\subsection{LSHR2 Level Set Sampling for Cauchy-Normal Model}

Our LSHR2 level set sampling methodology was implemented on the same Cauchy-normal model.  We start from the prior mode and move outwards in convex level sets, while running an exponentially-tilted hit-and-run algorithm within each level set in order to get samples from the posterior distribution.   As outlined in Section~\ref{exponentialtiltedwarmsampler}, each convex level set is defined by a threshold on the density that is slowly decreased in order to assure that each level set has a warm start. 

\begin{figure}[ht!]
\caption{Number of thresholds needed for our level-set hit-and-run sampler on the posterior density of the Cauchy-normal model as function of dimension.}
\label{cauchynormalnumthres}

\vspace{-1cm}

\begin{center}
\includegraphics[width=4in]{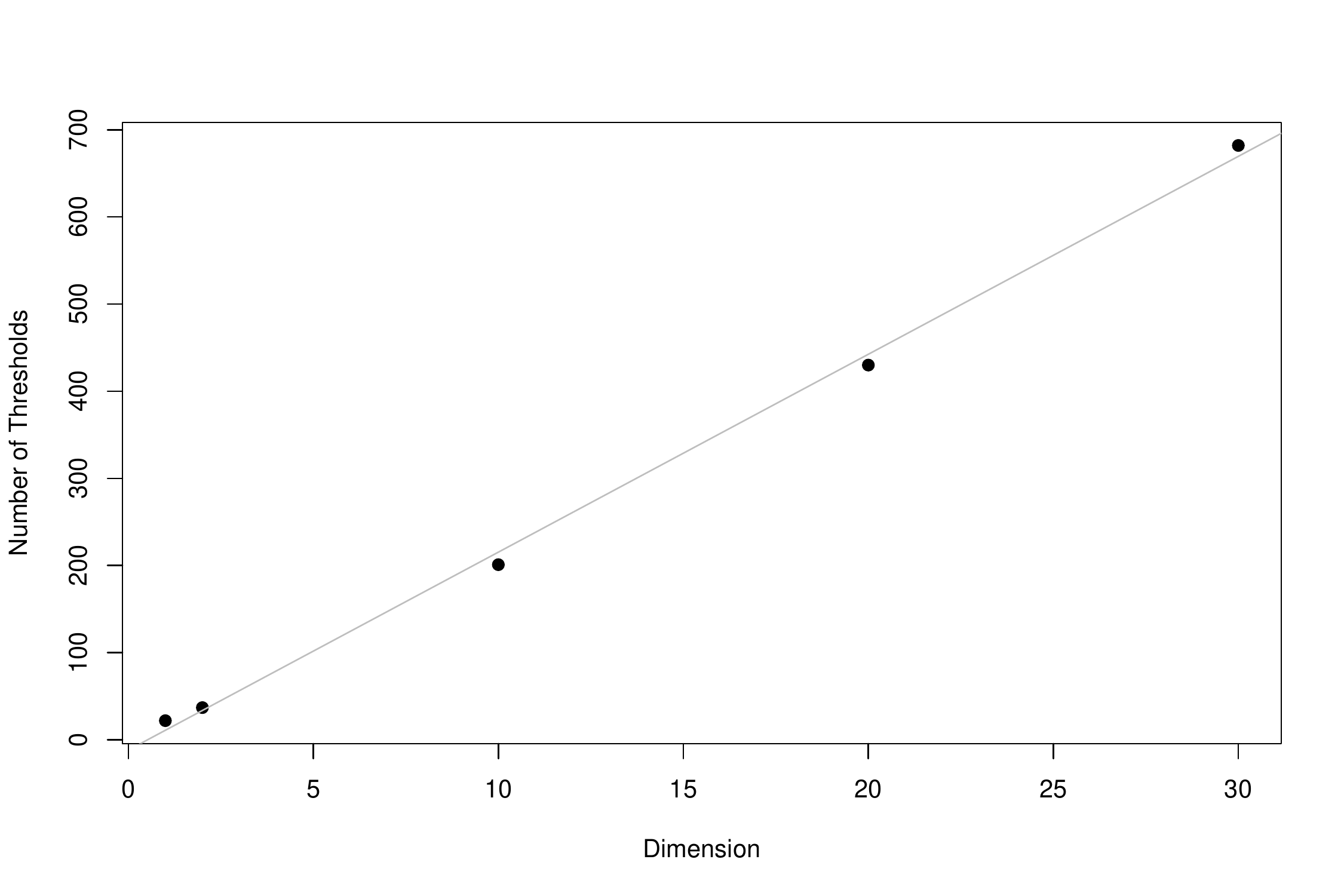}
\end{center}
\end{figure}

Figure~\ref{cauchynormalnumthres} shows the number of thresholds needed to fully explore the posterior distribution of the Cauchy-normal model for different number of dimensions.  Even for high dimensions ($d=10$ and $d=20$), we see that the posterior density is explored with a reasonable number of thresholds, and the approximately linear trend suggests that our algorithm would scale well to even higher dimensions.  

\begin{figure}[ht!]
\caption{Posterior samples of the first dimension $\theta_1$ of $\btheta$ from LSHR2 sampler for Cauchy-normal model in different dimensions $d$. Red curve in each plot represents the true posterior density.}
\label{cauchynormallevelsethists}

\vspace{-0.5cm}

\begin{center}
\includegraphics[width=6in]{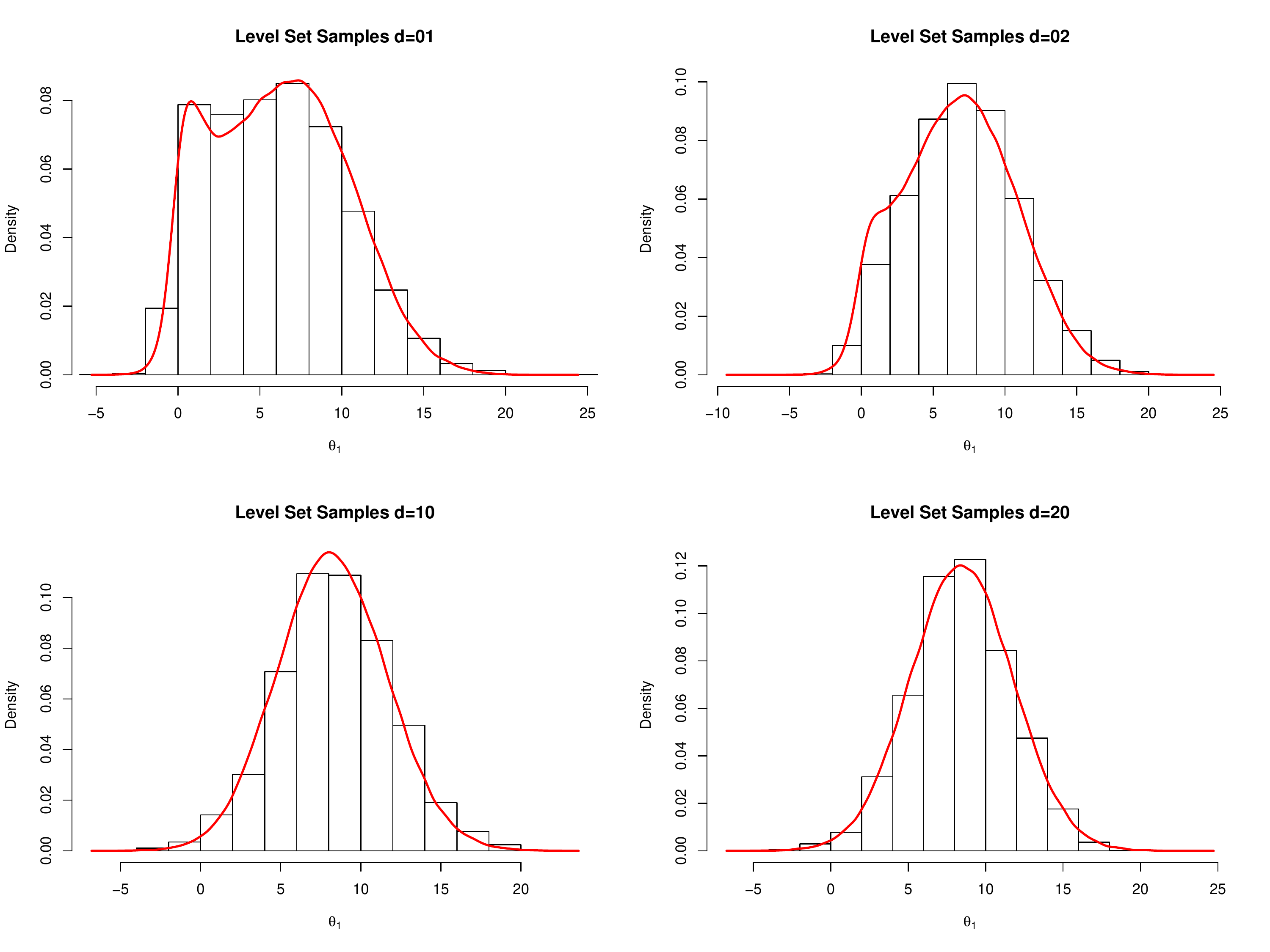}
\end{center}
\line(1,0){430}
\end{figure}

In Figure~\ref{cauchynormallevelsethists}, we compare the posterior samples for the first dimension $\theta_1$ from our LSHR2 sampler to the true posterior distribution for $d = 1, 2, 10$ and 20.   
Our LSHR2 level set sampler (with exponential tilting) samples closely match the true posterior density in both low and high-dimensional cases.   Comparing Figures~\ref{cauchynormalgibbshists} and Figures~\ref{cauchynormallevelsethists} clearly suggests that our level set hit-and-run methodology gives more accurate samples from the Cauchy-normal model in dimensions higher than one.

\section{Discussion}
In this paper, we have developed a general sampling strategy based on dividing a density into a series of level sets, and running the hit-and-run algorithm within each level set.    Our basic level-set hit-and-run sampler (LSHR1) can be applied to the broad class of pseudo-concave densities, which includes most distributions used in applied statistical modeling.   We illustrate our LSHR1 sampler on spike-and-slab density (Section~\ref{empirical1}), where our procedure performs much better in high dimensions than the usual Gibbs sampler.  

We also extend our sampling methodology to an exponentially-tilted level-set hit-and-run sampler (LSHR2).  Our LSHR2 algorithm can be applied to log-concave densities, which we illustrate with a simple multivariate normal example in Section~\ref{empirical2a}.  In this setting, our LSHR2 algorithm is more effective than Gibbs sampling when there is high dependence between parameters.  

Our LSHR2 can also be applied to exponentially-tilted quasi-concave densities, which arise frequently in Bayesian models as the posterior distribution formed from the combination of a log-concave likelihood and quasi-concave prior density.   We illustrate our LSHR2 sampler on the posterior density from a Cauchy normal model (Section~\ref{empirical2}), which can exhibit multi-modality under certain parameter settings.   Our procedure is able to sample effectively from this multi-modal posterior density even in high dimensions, where the Gibbs sampler performs poorly.  



\singlespacing

\begin{appendix}

\newpage

\section{The LSHR1 Algorithm} \label{warmsamplerdetails}

In this appendix, we give pseudo-code for our LSHR1 level-set hit-and-run sampler described in Section~\ref{warmsampleroutline}.  We must pre-specify a minimum value $K$ of the probability density $f(\cdot)$ as our stopping threshold.   The first threshold $t_1$ must be pre-specified, and we arbitrarily use $t_1 = 0.95*f(\bx_{\max})$.   Finally, the number of hit-and-run iterations per level set, $m$, must be chosen.   In our spike-and-slab example, we set the number of hit-and-run iterations per level set to $m = 1000$ based on the evaluation presented in our supplementary materials.    

\begin{algorithm}[ht]
\caption{LSHR1: Level-set Hit-and-run Sampler}
\begin{algorithmic}[1]
{\small
\State initialize $\bx = \bx_{\max}$ and $\Sigma_1 = {\rm I}$
\State initialize $t_1$ and set  $k = 1$
\State define current level set $C_1$ as $\{\bx : f(\bx) > t_1\}$
\For{$m$ iterations}
\State sample random direction $\bd$
\State calculate boundaries $\ba$ and $\bb$ of $C_1$ along direction $\bd$
\State sample new $\bx$ uniformly between $\ba$ and $\bb$ 
\State store all samples $\bx$ in $\{ \bx \}_1$
\EndFor
\While{$t_k > K$}
\State propose new $t_{\rm prop} < t_k$
\State define proposed level set $C_{\rm prop}$ as $\{\bx : f(\bx) > t_{\rm prop} \}$
\For{$m$ iterations}
\State sample random direction $\bd^\star$ and set $\bd = \Sigma_k^{1/2} \bd^\star$
\State calculate boundaries $\ba$ and $\bb$ of $C_1$ along direction $\bd$
\State sample new $\bx$ uniformly between $\ba$ and $\bb$ 
\If{$f(\bx) > t_{k}$}{$\, \hat{R}_{k:{\rm prop}} =  \hat{R}_{k:{\rm prop}} + 1$}\EndIf
\EndFor
\State $\hat{R}_{k:{\rm prop}} =  \hat{R}_{k:{\rm prop}}/m$
\If{$0.55 \leq \hat{R}_{1:{\rm prop}} \leq 0.8$}
\State $k = k + 1$
\State $t_k = t_{\rm prop}$; $C_{\rm k} = C_{\rm prop}$; $\hat{R}_{k:k+1} =  \hat{R}_{k:{\rm prop}}$; $\Sigma_{k+1} = {\rm Cov}(\bx)$
\State store all samples $\bx$ in $\{ \bx \}_k$
\EndIf
\EndWhile
\State define $n$ = number of level sets 
\For{$i=1,\ldots,n$}
\State $q_i = (t_{i-1}-t_{i}) \times \prod_{k=i}^n \hat{R}_{k:k+1}$
\State where $t_{0}$ is the maximum of $f(\cdot)$ and $\hat{R}_{n:n+1} = 1$
\EndFor
\For{$l=1,\ldots,L$}
\State Sample level set $s$ w.p.  $p_s = q_s / \sum q_s$
\State Sample point $\bx_l$ randomly from level set collection $\{ \bx \}_s$
\EndFor
}
\end{algorithmic}
\end{algorithm}
When considering the proposal for level set $k+1$, the initial $t_{\rm prop}$ is set to $t_{k} - (t_{k-1}-t_{k})$, so that our proposal moves the same distance in $t$ as the previous successful move.   If this initial $t_{\rm prop}$ is rejected for not being warm enough, we choose a new proposal $t_{\rm prop}$ as the midpoint of the old proposal and $t_{k}$, and so on. 

\newpage 

\section{The LSHR2 Algorithm} \label{expwarmsamplerdetails}

In this appendix, we give pseudo-code for LSHR2, our exponentially-weighted level-set hit-and-run sampler procedure described in Section~\ref{exponentialtiltedwarmsampler}.  We must pre-specify a minimum value $K$ of the probability density $f(\cdot)$ as our stopping threshold.   The first threshold $t_1$ must be pre-specified, and we arbitrarily use $t_1 = 0.95*f(\btheta_{\max})$ where $\btheta_{\max}$ is the mode of $f(\btheta)$.   We set the number of hit-and-run iterations per level set to $m = 1000$.

\begin{algorithm}[ht]
\caption{LSHR2: Exponentially-tilted Level-set Hit-and-run Sampler}
\begin{algorithmic}[1]
{\small
\State initialize $\btheta = \btheta_{\max}$ and $\Sigma_1 = {\rm I}$ 
\State initialize $p = \log g(\btheta_{\max})$ and $\btheta^\star = (\btheta,p)$
\State initialize $t_1$ and set  $k = 1$
\State define current level set $D_1$ as $ \{ \btheta^\star: f(\btheta) \geq t_1 \, , \, p < \log g(\by | \btheta) \} $
\For{$m$ iterations}
\State sample random direction $\bd$ in ($d+1$) dimensions
\State calculate boundaries $\ba$ and $\bb$ of $D_1$ along direction $\bd$
\State sample new $\btheta^\star$ proportional to $\exp(p)$ from line segment between $\ba$ and $\bb$
\State store all samples $\btheta^\star$ in $\{ \btheta^\star \}_1$
\EndFor
\While{$t_k > K$}
\State propose new $t_{\rm prop} < t_k$
\State define proposed level set $D_{\rm prop}$ as $ \{ \btheta^\star: f(\btheta) \geq t_{\rm prop} \, , \, p < \log g(\by | \btheta) \} $
\For{$m$ iterations}
\State sample random direction $\bd^\star$ in ($d+1$) dimensions and set $\bd = \Sigma_k^{1/2} \bd^\star$
\State calculate boundaries $\ba$ and $\bb$ of $D_{\rm prop}$ along direction $\bd$
\State sample new $\btheta^\star$ proportional to $\exp(p)$ from line segment between $\ba$ and $\bb$
\If{new $\btheta^\star \in D_k$}{$\, \hat{R}_{k:{\rm prop}} =  \hat{R}_{k:{\rm prop}} + 1$}\EndIf
\EndFor
\State $\hat{R}_{k:{\rm prop}} =  \hat{R}_{k:{\rm prop}}/m$
\If{$0.55 \leq \hat{R}_{1:{\rm prop}} \leq 0.8$}
\State $k = k + 1$
\State $t_k = t_{\rm prop}$; $D_{\rm k} = D_{\rm prop}$; $\hat{R}_{k:k+1} =  \hat{R}_{k:{\rm prop}}$; $\Sigma_{k+1} = {\rm Cov}(\bx)$
\State store all samples $\btheta^\star$ in $\{ \btheta^\star \}_k$
\EndIf
\EndWhile
\State define $n$ = number of level sets 
\For{$i=1,\ldots,n$}
\State $q_i = (t_{i-1}-t_{i}) \times \prod_{k=i}^n \hat{R}_{k:k+1}$
\State where $t_{0}$ is the maximum of $f(\cdot)$ and $\hat{R}_{n:n+1} = 1$
\EndFor
\For{$l=1,\ldots,L$}
\State Sample level set $s$ w.p.  $p_s = q_s / \sum q_s$
\State Sample point $\btheta^\star_l$ randomly from level set collection $\{ \btheta^\star \}_s$
\EndFor
}
\end{algorithmic}
\end{algorithm}
When considering the proposal for level set $k+1$, the initial $t_{\rm prop}$ is set to $t_{k} - (t_{k-1}-t_{k})$, so that our proposal moves the same distance in $t$ as the previous successful move.   If this initial $t_{\rm prop}$ is rejected for not being warm enough, we choose a new proposal $t_{\rm prop}$ as the midpoint of the old proposal and $t_{k}$, and so on.

\section{Gibbs Switching Calculation} \label{gibbsswitch}

For a variable in the zero component, with norm $||{\bf
x}||_2^2 \approx
d \sigma_0^2 $, the probability of a switch is  
\begin{eqnarray*}
P\left(I = 1 \, | \, ||{\bf x}||_2^2  \approx
d \sigma_0^2\right) 
& = &  \frac{\sigma_1^{-d} \exp(- ||{\bf x}||_2^2 / 2\sigma_1^2)}{\sigma_1^{-d} \exp(- ||{\bf x}||_2^2 / 2\sigma_1^2) + \sigma_0^{-d} \exp(- ||{\bf x}||_2^2 / 2\sigma_0^2)} \\
& = &  \frac{\sigma_1^{-d} \exp(- d \sigma_0^2 / 2\sigma_1^2)}{\sigma_1^{-d} \exp(- d \sigma_0^2 / 2\sigma_1^2) + \sigma_0^{-d} \exp(- d / 2)} \\
& = & \frac{\left(\frac{\sigma_0}{\sigma_1}\right)^d \exp \left(\frac{d}{2}\left(1 - \frac{\sigma_0^2}{\sigma_1^2}\right)\right)}{\left(\frac{\sigma_0}{\sigma_1}\right)^d \exp \left(\frac{d}{2}\left(1 - \frac{\sigma_0^2}{\sigma_1^2}\right)\right) + 1}\\
& \approx & \frac{\left(\frac{\sigma_0}{\sigma_1}\right)^d \exp \left(\frac{d}{2}\right)}{\left(\frac{\sigma_0}{\sigma_1}\right)^d \exp \left(\frac{d}{2}\right) + 1}\\
\end{eqnarray*}
since $\sigma_0 \ll \sigma_1$.    Now, as $d$ increases, $(\sigma_0/\sigma_1)^d$ goes to zero, and so 
\begin{eqnarray*}
 \frac{\left(\frac{\sigma_0}{\sigma_1}\right)^d \exp \left(\frac{d}{2}\right)}{\left(\frac{\sigma_0}{\sigma_1}\right)^d \exp \left(\frac{d}{2}\right) + 1}  \approx  \left(\frac{\sigma_0}{\sigma_1}\right)^d \exp \left(\frac{d}{2}\right) = \left(\frac{\sigma_0\sqrt{e}}{\sigma_1}\right)^d 
\end{eqnarray*}

\end{appendix}

\begin{footnotesize}
\bibliography{references}
\end{footnotesize}

\end{document}